\documentclass[twocolumn]{aastex631}

\received{Aug 30, 2024}

\begin{document}

\title{The ALMA Survey of Gas Evolution of PROtoplanetary Disks (AGE-PRO): \\XII. Extreme millimetre variability detected in a Class II disc}



\author[0000-0002-1575-680X]{James M. Miley}
\affiliation{Departamento de Física, Universidad de Santiago de Chile, Av. Victor Jara 3659, Santiago, Chile}
\affiliation{Millennium Nucleus on Young Exoplanets and their Moons (YEMS), Chile }
\affiliation{Center for Interdisciplinary Research in Astrophysics Space Exploration (CIRAS), Universidad de Santiago de Chile, Chile}

\author[0000-0002-1199-9564]{Laura M. P\'erez}
\affiliation{Departamento de Astronom\'ia, Universidad de Chile, Camino El Observatorio 1515, Las Condes, Santiago, Chile}

\author[0000-0002-7238-2306]{Carolina Agurto-Gangas}
\affiliation{Departamento de Astronom\'ia, Universidad de Chile, Camino El Observatorio 1515, Las Condes, Santiago, Chile}

\author[0000-0002-5991-8073]{Anibal Sierra}
\affiliation{Departamento de Astronom\'ia, Universidad de Chile, Camino El Observatorio 1515, Las Condes, Santiago, Chile}
\affiliation{Mullard Space Science Laboratory, University College London, Holmbury St Mary, Dorking, Surrey RH5 6NT, UK}

\author[0000-0002-8623-9703]{Leon Trapman}
\affiliation{Department of Astronomy, University of Wisconsin-Madison, 475 N Charter St, Madison, WI 53706, USA}

\author[0000-0002-4147-3846]{Miguel Vioque}
\affiliation{European Southern Observatory, Karl-Schwarzschild-Str. 2, 85748 Garching bei München, Germany}
\affiliation{Joint ALMA Observatory, Avenida Alonso de Córdova 3107, Vitacura, Santiago, Chile}

\author[0000-0002-2358-4796]{Nicolas Kurtovic}
\affiliation{Max-Planck-Institut fur Astronomie (MPIA), Konigstuhl 17, 69117 Heidelberg, Germany}

\author[0000-0001-8764-1780]{Paola Pinilla}
\affiliation{Mullard Space Science Laboratory, University College London, Holmbury St Mary, Dorking, Surrey RH5 6NT, UK}

\author[0000-0001-7962-1683]{Ilaria Pascucci}
\affiliation{Lunar and Planetary Laboratory, The University of Arizona, Tucson, AZ 85721, USA}

\author[0000-0002-0661-7517]{Ke Zhang}
\affiliation{Department of Astronomy, University of Wisconsin-Madison, 475 N Charter St, Madison, WI 53706, USA}

\author[0009-0004-8091-5055]{Rossella Anania}
\affiliation{Dipartimento di Fisica, Università degli Studi di Milano, Via Celoria 16, I-20133 Milano, Italy}

\author[0000-0003-2251-0602]{John Carpenter}
\affiliation{Joint ALMA Observatory, Avenida Alonso de Córdova 3107, Vitacura, Santiago, Chile}

\author[0000-0002-2828-1153]{Lucas A. Cieza}
\affiliation{Instituto de Estudios Astrofísicos, Universidad Diego Portales, Av. Ejercito 441, Santiago, Chile}
\affiliation{Millennium Nucleus on Young Exoplanets and their Moons (YEMS), Chile }

\author[0000-0003-0777-7392]{Dingshan Deng}
\affiliation{Lunar and Planetary Laboratory, The University of Arizona, Tucson, AZ 85721, USA}

\author[0000-0003-4907-189X]{Camilo Gonz\'alez-Ruilova}
\affiliation{Instituto de Estudios Astrofísicos, Universidad Diego Portales, Av. Ejercito 441, Santiago, Chile}
\affiliation{Millennium Nucleus on Young Exoplanets and their Moons (YEMS), Chile }
\affiliation{Center for Interdisciplinary Research in Astrophysics Space Exploration (CIRAS), Universidad de Santiago de Chile, Chile}

\author[0000-0003-4853-5736]{Giovanni P. Rosotti}
\affiliation{Dipartimento di Fisica, Università degli Studi di Milano, Via Celoria 16, I-20133 Milano, Italy}

\author[0000-0003-3573-8163]{Dary A. Ruiz-Rodriguez}
\affiliation{National Radio Astronomy Observatory; 520 Edgemont Rd., Charlottesville, VA 22903, USA}
\affiliation{Joint ALMA Observatory, Avenida Alonso de Córdova 3107, Vitacura, Santiago, Chile}

\author[0000-0001-9961-8203]{Estephani E. TorresVillanueva}
\affiliation{Department of Astronomy, University of Wisconsin-Madison, 475 N Charter St, Madison, WI 53706, USA}

\begin{abstract}

Variability of millimetre wavelength continuum emission from Class II protoplanetary disks is extremely rare, and when detected it is usually interpreted as originating from non-thermal emission mechanisms that relate to the host star itself rather than its disk. During observations made as part of the AGE-PRO ALMA Large Program, significant variability in the brightness of the 2MASS J16202863-2442087 system was detected between individual executions.
We report the observed properties of the variability detected at millimetre wavelengths and investigate potential driving mechanisms.
To investigate the nature of the variability we construct a light curve from the continuum observations and analyse images constructed from both flaring and quiescent emission. We characterise the dust disk around the star through analysis in the image and visibility plane, and carry out kinematic analysis of CO (2--1) emission from the gas disk.
The continuum flux decays by a factor of 8 in less than an hour, and by a factor of 13 within 8 days. The peak brightness coincides with an expected brightness maximum extrapolated from the periodicity of previously observed optical variability.

The flare is most likely the product of synchrotron emission in the close vicinity of the star. The nature of the millimetre flare closely resembles those detected in very close binary systems, and may be due to the interaction of magnetic fields in an as yet undetected binary. Alternatively, if the central host is a single-star object, the flare may be due to the interaction of magnetic field loops at the stellar surface or a strong accretion burst.

\end{abstract}


\section{Introduction}

Pre-main sequence stars hosting circumstellar disks exhibit variable emission across a range of magnitudes, frequencies and time scales. A review by \citet{Fischer2023AccretionAssembly} identifies broad categories of the most common sources of variability which includes accretion related events, effects due to the circumstellar-disks, extinction related events, stellar phenomena and binary-related phenomena. 
Extreme variability due to accretion driven outbursts is quite common at early embedded ($\sim$ Class I) stages \citep{Hartmann1985OnObjects, Herbst1994CatalogueVariability}, for which there can be significant repercussions for planet formation within the circumstellar disk \citep[e.g.][]{Cieza2016ImagingOutburst}. Extinction related variability is important when considering Class II systems, as seen in so-called dipper stars at optical wavelengths \citep[e.g.][]{Ansdell2016YOUNGK2,Sissa2019TheDisk}, where circumstellar disks occult stellar light. A less well understood type of variability, and one that is rarely caught by observations, is variability in the millimetre (mm) wavelength regime.

Typically, the main source of mm/sub-mm emission from disks originates from cold gas and relatively large ($\sim$mm sized) dust grains, which are expected to be produce quiescent emission.
Transient flaring of mm emission by a factor of a few over short timescales (i.e. timescales of hours or days) is extremely rare, and is associated with non-thermal emission related to the interaction of magnetic field lines. \citet{Massi2006SynchrotronA} discuss indicative three scenarios in which this can take place, each of which varies in their flare location and in the geometrical structure of the magnetic fields involved. The first scenario is in a single star system, where magnetic loops can be created through the dynamo theory used to explain sunspots \citep{Parker1955THEFIELD}. Differential rotation and convection within the stellar interior create magnetic loops on the stellar surface and the interaction between two loops results in the acceleration of electrons to very high energies until they escape along field lines. Those that escape towaxrds the corona are responsible for creating radio bursts \citep{Heryvaerts1977ANPHENOMENON}. Alternatively, one can consider young systems in which there remains a circumstellar disk. Threading of the stellar field lines within the inner disk can lead to excitation of electrons and subsequent flaring if these field lines become stretched or distorted due to rapid rotation of the star \citep{Massi2008InteractingStreamers}. The final scenario occurs in close binary systems, in particular those of later spectral type, because in these stars the convective region of the stellar interior is at the surface and stellar rotation is fast leading to higher magnetic activity. If magnetic loops from each of the companions interact at a moment of close-passage in the orbit, then this can also result in radio flares. 

The disparate observational setups and temporal sampling cadence used to observe these various variability events, which are often discovered serendipitously, complicate the characterisation and comparison of such phenomena. The measured amplitudes of mm flux increases can range from a factor of a few, to variations of orders of magnitude depending on the timescale over which they are measured. Further uncertainty is introduced when considering that the calibrator sources used to calibrate flux in the observations vary themselves, therefore the accuracy of any given flux is dependent upon how well the calibrators are monitored \citep[see][where this is investigated for the specific case of ALMA observations]{Francis2020OnWindows}. Despite these challenges, data currently in hand shows that variability has been observed to occur on timescales of days, hours, and in some cases flux increases that only last for minutes.


This work presents a detection of an extreme flare of mm emission measured towards the star 2MASS J16202863-2442087 (labelled UppSco~7 in AGE-PRO survey papers, and referred to as J16202 hereafter). The star is an M2 spectral type \citep{Esplin2018Association} situated $\sim$153~pc from Earth \citep{GaiaCollaboration2018VizieR2018}. We characterise the variability of the mm continuum emission as detected by AGE-PRO observations, and use analysis of gas and dust disk emission in order to contextualise the discussion of possible  origin mechanisms for the increase in mm emission, with reference to previously published systems with variable mm continuum flux. 


\section{Observations}

\begin{figure}
    \centering
    \includegraphics[width=1.0\linewidth]{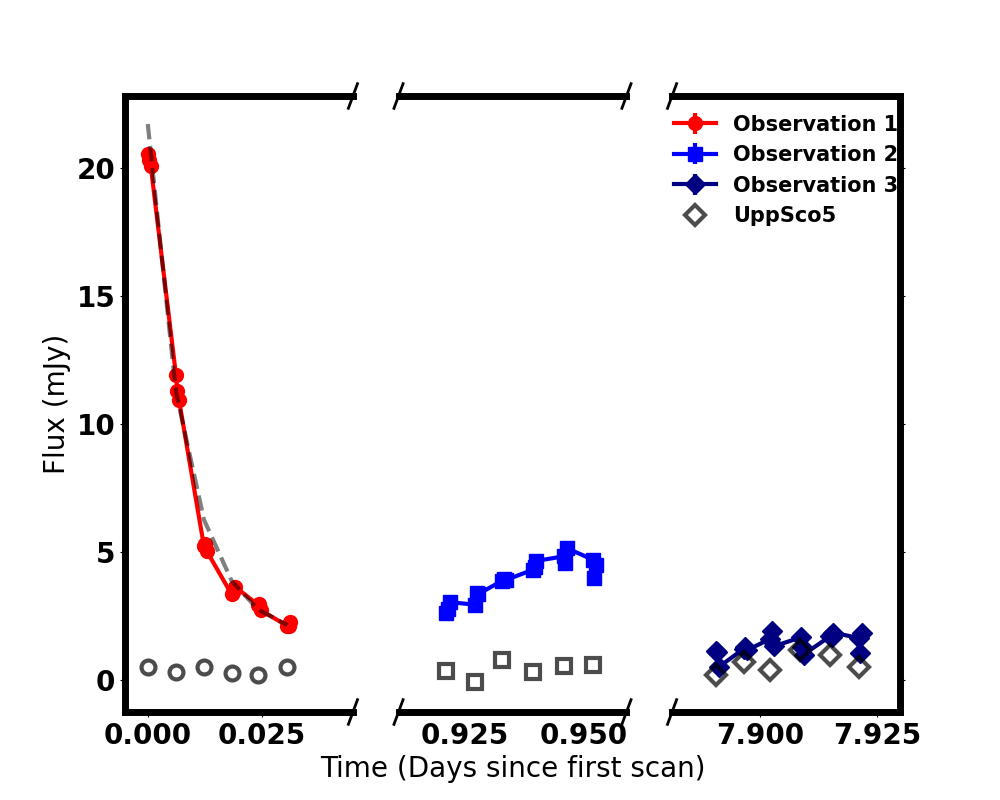}
    \caption{Flux measurements from each scan made in the short-baseline observations towards J16202. In this plot each scan is divided into three equally sized time bins, and the flux is measured by fitting a 2D Gaussian to the unresolved continuum emission. The black points show similar measurements made for the source UppSco~5, which was also observed as part of these executions.}
    \label{fig:mm_lightcurve}
\end{figure}

\begin{figure*}[t!]
    \centering
    \includegraphics[width=1\linewidth]{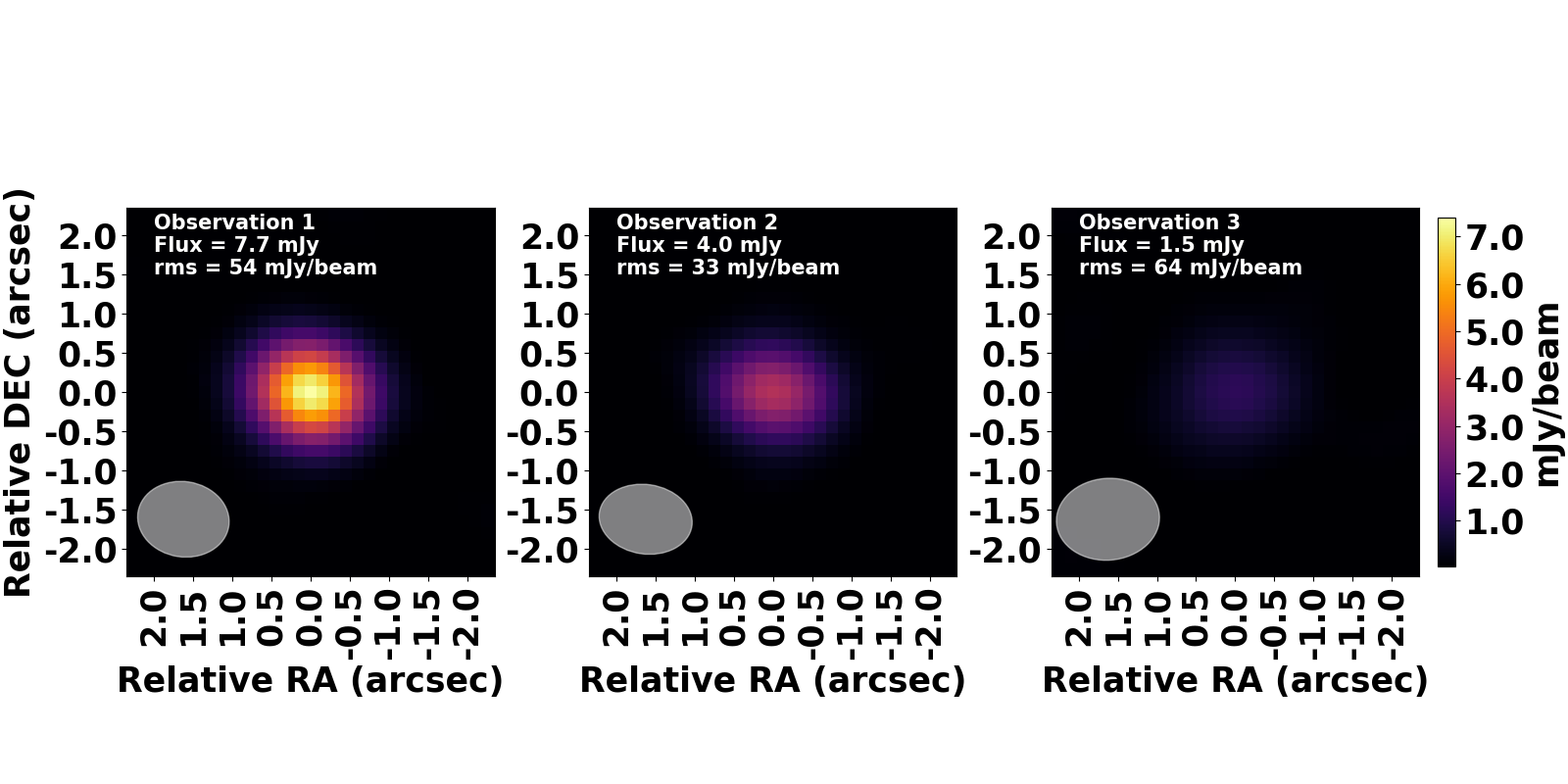}
    \caption{Images constructed from the three short baseline observations taken at 1.3~mm of J16202 that demonstrate how flux from the disk varies substantially. All three images use the same colourmap as described by the colourbar. The synthesised beam of each observation is shown in the bottom left of each image, annotated in the top left hand corner are the integrated flux over the course of the observation, and the measured rms of each image.}
    \label{fig:cont_SB_ims}
\end{figure*}

ALMA observations of the Class II protoplanetary disk around J16202 were taken as part of the AGE-PRO large program which included two observing setups, one in Band 6 and one in Band 7. Full details of the correlator setup and the molecular lines targeted by these observations can be found in \citet{Zhang2024TheResults} and \citet{Agurto-Gangas2024TheRegion}. In this work we focus on the observations most important for constraining the flare in mm emission observed in J16202, specifically the Band 6 short-baseline (SB) observations from which we study the continuum emission and $^{12}$CO~(J=2--1) molecular line emission. J16202 belongs to the Upper Scorpius region \citep{Luhman2022AComplex} from which 10 objects were studied by the AGE-PRO ALMA large program as detailed in \citet{Agurto-Gangas2024TheRegion}. Each of these objects was observed with a combination of a compact and a more extended antenna configuration, a summary of which is provided in Table \ref{tab:obsList}. The flare in mm emission was detected during compact configuration observations of J16202 on 30th March 2023 (labelled henceforth as Observation 1), two subsequent SB observations were made on the 31st March and 7th April (labelled as Observation 2 and Observation 3 respectively). J16202 was part of an observing block including 4 other protoplanetary discs, which are observed consecutively and share the same calibrators. These observations were approximately 50 minutes in length including overheads, in which the science targets were observed in 6 individual scans of around 90 seconds each. Calibration observations were observed in between these on-target scans. We see no evidence of significant atmospheric line variability between the observations. Of the observations made on three separate occasions with the compact configuration, the second execution was discarded from the full AGE-PRO analysis due to semipass quality, but we include the data here to add data points to the light curve and investigate the evolution of the measured flux from J16202. In July of 2022 observations were made using an antennas configuration with longer baselines (LB), none of these later observations detected flaring emission. For completeness, the LB observations are also described in Table \ref{tab:obsList} and they are included when constructing continuum images only where indicated.

\begin{table*}[]
\centering
\begin{tabular}{c|c|c|c|c|c}
\textbf{Label}& \textbf{Observation Start} & \textbf{ALMA config.} & \textbf{Baseline range} & \textbf{pwv (mm)} & \textbf{Calibrators} \\ \hline
Observation 1&2022-03-30 8:27 & C43-2 & 14m -- 313m & 0.9 & J1517-2422, J1625-2527 \\ \hline
Observation 2&2022-03-31 6:48 & C43-2 & 14m -- 330m & 1.0 & J1427-4206, J1625-2527 \\ \hline 
Observation 3&2022-04-07 5:49 & C43-2 & 14m -- 313m & 2.6 & J1517-2422, J1625-2527 \\ \hline
&2022-07-03 23:08 & C43-5 & 15m -- 1301m & 1.1 & J1517-2422, J1626-2951 \\ \hline
&2022-07-04 1:05 & C43-5 & 15m -- 1301m & 1.1 & J1427-4206, J1626-2951 \\ \hline
&2022-07-04 2:42 & C43-5 & 15m -- 1301m & 1.2 & J1517-2422, J1626-2951 \\ \hline
&2022-07-05 0:39 & C43-5 & 15m -- 1996m & 1.8 & J1427-4206, J1626-2951 \\ \hline
&2022-07-17 23:38 & C43-5 & 15m -- 2617m & 0.4 & J1427-4206, J1700-2610 
\end{tabular}
\caption{AGE-PRO log of Band 6 ALMA observations towards J16202.}
\label{tab:obsList}
\end{table*}

\section{Results}
\label{sec:results}

\subsection{Detection of mm flare emission}

The flux measured towards J16202 was found to vary between the three executions made with the compact configuration. ALMA observations are comprised of scans that alternate between the science target and calibrators. Each observation of a science target is therefore comprised of multiple short scans. When the observed data of J16202 is separated into individual scans, a varying light-curve emerges as is shown in Figure \ref{fig:mm_lightcurve}, while the millimetre continuum images from each individual day of observation are shown in Figure \ref{fig:cont_SB_ims}. The flux is initially measured at 20.5~mJy during the first scan of Observation 1, but by the end of the execution the measured flux decreased to a level of $\approx$2.5~mJy, a factor of 8.2 times fainter than at the beginning of the observation.

The largest flux measurement is taken during the first scan of Observation 1, and so from this data alone we cannot confidently constrain the true peak flux that occurred during this flare. None of the other 4 objects observed as part of the same observing block show any variation in their continuum emission and neither do any of the calibrators observed alongside the science targets. Observation 2 shows an increase in flux of a factor 2 compared to quiescent levels, which is also a significant increase in continuum emission. This is reminiscent of the active quiescent phases identified by \citet{Lovell2024SMA283572}. Observation 3 is the best characterisation of the quiescent state in short baseline observations, in which we measure an approximately constant flux of $\approx$1.5~mJy, which is a factor of 13.7 less than the peak measured flux from Observation 1. The long-baseline (LB) observations taken a few months later show a consistent flux of $\approx$ 0.89~mJy, although we must note that this is not a direct comparison of the flux from the disk measured with the short-baseline observations due to the different antenna configuration utilised. It may be that the flux of J16202 has continued to dim between Observation 3 and the LB  executions, or that the different ALMA configuration used may have resulted in filtering out of more extended emission to which the more extended configuration is less sensitive. The maximum resolvable scale of the extended configuration observations can be estimated by considering the 5th percentile baseline length (97.9m) and the observing wavelength which gives size scale of 2.8 arcsec. With these precautions in assessing the measurements in mind, the flux values from the LB observations are nevertheless clearly more consistent with that of quiescent Observation 3 than flaring Observations 1 or 2.


\begin{figure}
    \centering
    \includegraphics[width=\linewidth]{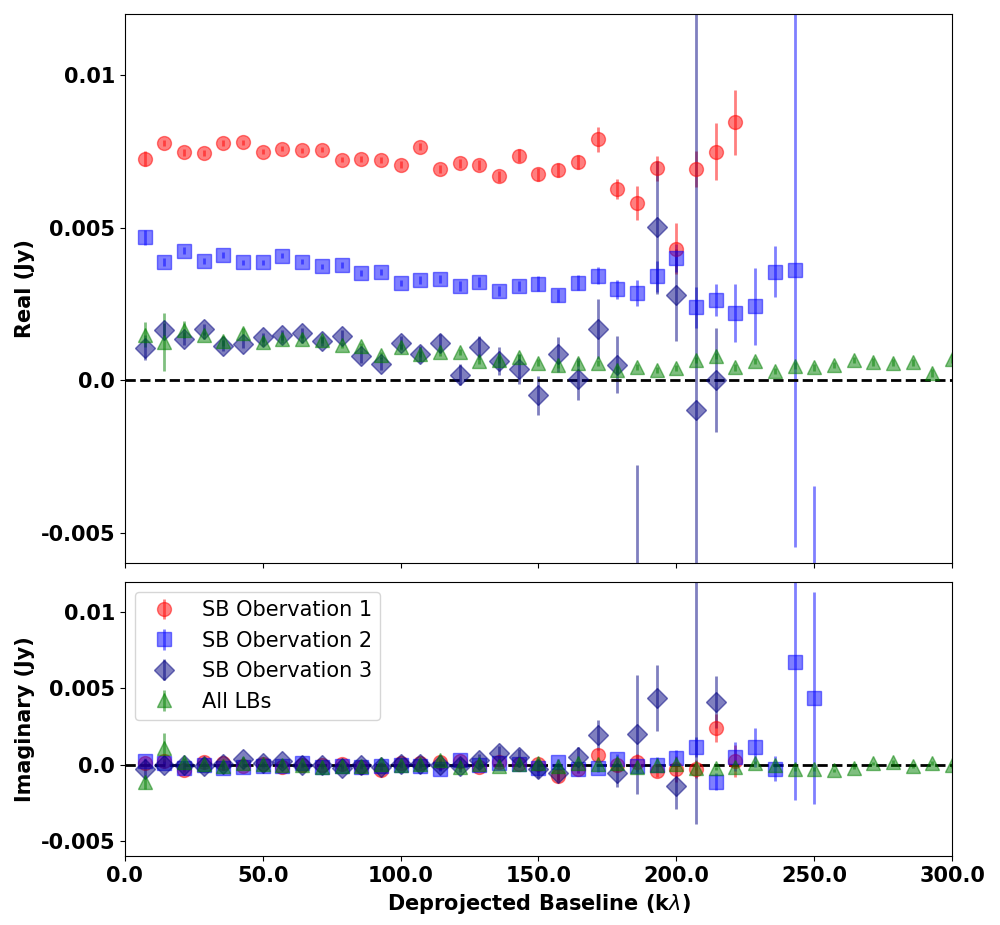}
    \caption{Observed visibilities plotted as a function of deprojected baseline length for individual short-baseline (SB) and combined long-baseline (LB) observations towards J16202.}
    \label{fig:plot_obs_vis}
\end{figure}

The continuum image used for AGE-PRO analysis does not include the flaring observations, because their inclusion masks the underlying brightness distribution. This has implications for image fidelity; the combination of compact and more extended configurations ensures better coverage of uv space, but removing two of the three compact-configuration observations reduces the coverage at small uv-distances which are sampled by the short-baselines. This cannot explain the dramatic difference in brightness that is measured however. To investigate further we may consider the observed visibilities, as presented in Figure \ref{fig:plot_obs_vis}. The observed visibilities are de-projected using inclination 32.7$^\circ$ and position angle 179.0$^\circ$ \citep{Agurto-Gangas2024TheRegion}, then plotted as a function of de-projected baseline length, as are the the concatenated visibilities of the long-baseline observations taken some months later. It is immediately clear from Figure \ref{fig:plot_obs_vis} that Observations 1 and 2 are tracing different emission from Observation 3 and the long baseline observations. Both of the presumed flaring observations are clearly displaced on the y axis and their visibilities do not converge towards 0 at long baseline lengths, indicating that the emission is not spatially resolved.
On the other hand, the real components of the de-projected visibilities of the long-baseline observations trace those of Observation 3 very closely, which suggests both are measuring emission from a quiescent state rather than the heightened emission seen in Observation 1 and 2.  An offset of the visibilities on the flux axis for all baseline lengths occurs when an object is unresolved. For the SB observations the maximum baseline was 313~m, meaning anything of size $\leq 0.86\arcsec$ will be unresolved. 

\begin{figure}
    \centering
    \includegraphics[width=\linewidth]{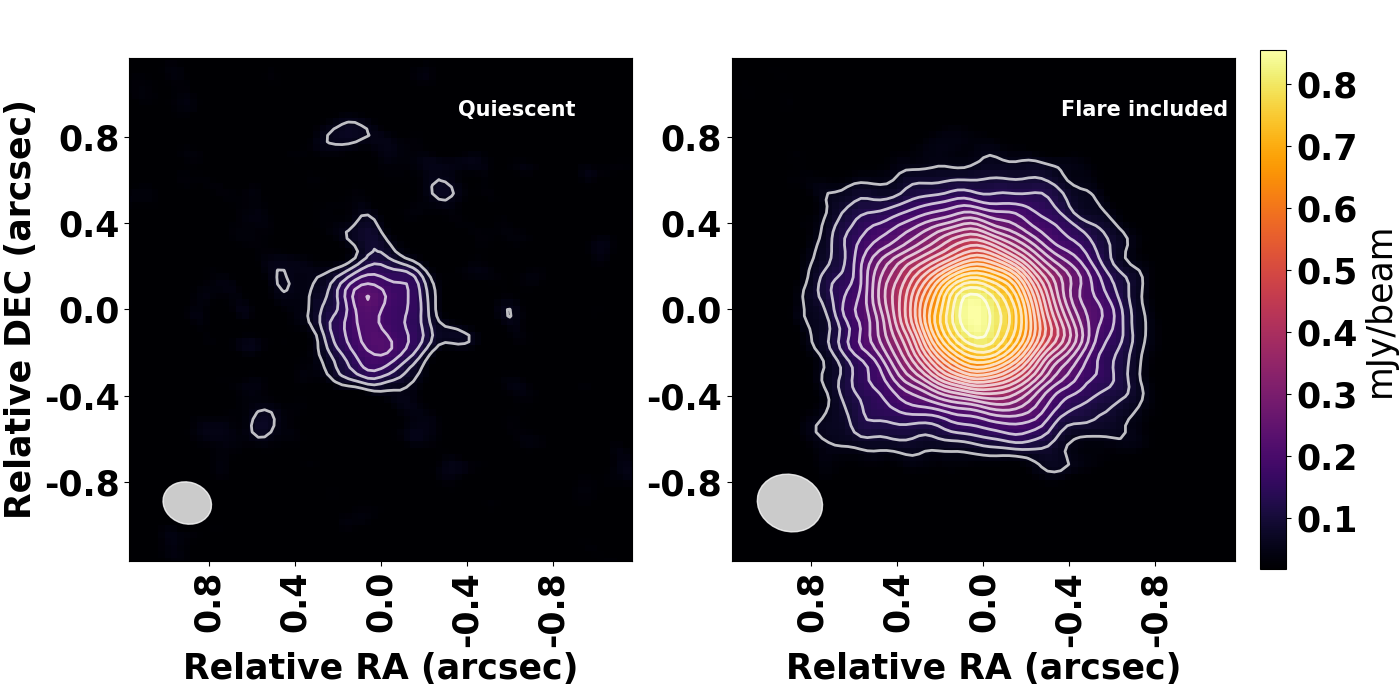}
    
    \caption{Continuum image of J16202 combining all observations made. In the left panel is the final image as presented in the AGE-PRO sample, constructed from long-baseline observations and Observation 3 using short-baselines. The right hand panel show the image when all executions are concatenated, including all long-baseline observations and all short-baseline observations. Contours begin at 3$\sigma$ and increase in steps of 2$\sigma$. The beam size achieved in the non-flare image is 0.228\arcsec$\times$0.194\arcsec, with flaring included the achieved beam size is 0.307\arcsec$\times$0.264\arcsec.}
    \label{fig:flare_cont_ims}
\end{figure}

While we will further discuss the origin and complexity of the brightening of the continuum emission in the next sections, for simplicity we will refer to this epoch as a `mm flare' throughout this work. 
Figure \ref{fig:flare_cont_ims} shows images created with and without including the flaring observations. In the left hand panel of Figure \ref{fig:flare_cont_ims} the peak of emission is not centred in the disk, instead at the disk centre we see a plateau in emission which might be indicating an unresolved inner cavity, or a decrease in density towards innermost regions. In the right hand panel of Figure \ref{fig:flare_cont_ims}, the bright emission from the flaring observations dominates the image, and a strong central peak is observed with no evidence of an inner cavity. Due to the brighter peak, the colour scale of the image in the right hand panel appears to suggest a larger disk size, but in fact the R$_{90}$ value for each of the discs is consistent within the error bars. A true difference in disk size is not detected here, the increase in brightness originates from a central, unresolved source (as shown in the visibilities presented in Figure \ref{fig:plot_obs_vis} and the radial flux profiles extracted from the image plane in Figure \ref{fig:uvspace}). In the following section we scrutinise these hints of a cavity by modelling observed visibilities.

Another indicative property of the flare is the spectral index, $\alpha$, where \( F \propto \nu^{-\alpha}  \). Blackbody emission has a characteristic $\alpha=2$.
In passively irradiated protoplanetary discs the spectral index depends upon dust composition, grain size and optical depth, and for disk-averaged values it is typically measured at $\approx$2.3 \citep{Andrews2020ObservationsStructures}. A steeper index can be interpreted as evidence of dust evolution, for example a collisional cascade that grinds particles to smaller sizes leads to $\alpha$ values closer to 3, as can be seen in debris discs for example \citep{Lohne2020RelatingWavelengths}. Optically thin, non-thermal brehmstrahlung radiation has $\alpha \leq$ -0.1, using the current sign convention. Ideally, the spectral index should be calculated over as wide a frequency range as possible, but as we lack simultaneous multi-wavelength data, here we calculate an intra-band $\alpha$ for J16202 using the two continuum dedicated windows at either end of the total bandwidth, with representative frequencies 218 and 234 GHz. We can scrutinise the flaring emission by comparing $\alpha$ measured in each of the three observations as well as in the combined dataset. These results are given in Table \ref{tab:specInd}.
During the flaring and active quiescent phases the non-thermal emission dominates emission and the spectral index is much shallower as a result, whereas in the quiescent phase, the spectral index is much steeper. We note that our intra-band spectral index is calculate with a relatively short lever-arm in frequency. \citet{Francis2020OnWindows} find that there is a flux calibration uncertainty between spectral windows from a single ALMA band of $\approx$0.8\%. These additional uncertainties are small however compared to the stark variation in $\alpha$ that is apparent in Table \ref{tab:specInd}. Ideally we require simultaneous (or as close to simultaneous as is possible), multi-frequency observations of the system with which to calculate a spectral index during both flaring and quiescent periods.


\begin{table}[]
\centering
\begin{tabular}{l|c}
Dataset & Spectral Index, $\alpha$ \\ \hline
Observation 1 (flare decay)& 0.1 $\pm$ 0.1 \\
Observation 2 (active quiescent) & -0.3 $\pm$ 0.1\\
Observation 3 (quiescent)& 3.4 $\pm$ 0.5 \\
All SBs Combined & 0.7 $\pm$ 0.1\\
Final data set (Obs. 3 + LBs)& 3.9 $\pm$ 0.6\\
\end{tabular}
\vspace{5pt}
\caption{Spectral indices counted for different portions of the observed data. The first row is the calculation for Observation 1, in which the strongest flaring is measured. The third row calculates $\alpha$ for Observation 3, the best representation of quiescent flux recorded by the short-baseline observations. The penultimate row gives the spectral index calculated when all short-baseline observations combined, whereas the final row measures the spectral index from the final dataset used in AGE-PRO analysis which combines Observation 3 with the LB observations. }
\label{tab:specInd}
\end{table}



Unnoticed variable mm flux in Class II discs could lead to inaccuracies in the calculation of  disk mass estimates derived from dust continuum observations. Assuming optically thin emission, the disk dust mass can be estimated using

\begin{equation}
    F_\nu = \frac{B_\nu(T) \kappa_\nu M_{\rm dust}}{d^2},
    \label{eqn:dustmass}
\end{equation}

where $B_\nu(T)$ is the Planck function evaluated at temperature T, $M_{\rm dust}$ is the total disk dust mass, $d$ is the distance to the star from Earth of 153.0~pc and $\kappa_\nu$ is the dust opacity. If we assume that all received emission is thermal, the disk mass estimate would increase proportionally with the measured flux. The integrated flux measured from the image including flaring observations (i.e. the right hand panel of Figure \ref{fig:flare_cont_ims}), is 3.5$\times$ greater than the integrated flux quiescent-only image, meaning the disk mass size would increase by the same factor. Equation \ref{eqn:dustmass}, and in particular the $\kappa_\nu$ term, carries large uncertainties that may lead to inaccuracies of between factors of a few to orders of magnitude \citep[see for example the disk mass estimates of TW Hya highlighted by][]{Miotello2022SettingProperties}. 

It is difficult to estimate the potential impact of ignoring continuum variability in dust mass constraints. Disentangling true flux variation from typical flux calibration uncertainties will be difficult where variability is of low levels. Amongst the most famous, well-studied protoplanetary disk systems that are regularly observed,  mm flux variation has not been detected at levels significant enough to cause concern. However such verification is lacking from the much larger population of generally fainter and more compact disks. Many of these systems may have been observed in  large surveys of star-forming regions that employ only a single pointing towards each target, in which case a factor of $\approx$3.5 such as we measure here can easily go unnoticed.

\subsection{Visibility analysis of the J16202 dust distribution}

We now investigate further the innermost regions of the system, where there is a hint of an inner cavity in the dust distribution in the left panel of Figure \ref{fig:flare_cont_ims}. For this analysis we fit to the final AGEPRO data set, which does not include observations that detect variable emission.
If the central object of J16202 is an as-yet-undetected inner binary, we would expect an inner cavity to be carved of size $\approx 3.5\times ~$a$_{\rm bin}$, where a$_{\rm bin}$ is the size of a circular binary orbit \citep{Ragusa2020TheDiscs}. Indeed, the cavity size is expected to be even greater than this in the case of an eccentric orbit \citep[e.g.][]{Kurtovic2022ThePlanet,Penzlin2024ViscousCavity}. An inner cavity can not be considered as conclusive proof of an inner binary, many so-called transition disks have shown cleared inner regions \citep{Francis2020Dust-depletedObservations}, but this is a particularly interesting property for us to constrain in the case of J16202 given that so many of the other extreme mm flares seen in Class II objects have come from close or even spectroscopic binary systems.

\begin{figure}
    \centering
    \includegraphics[width=1.03\linewidth]{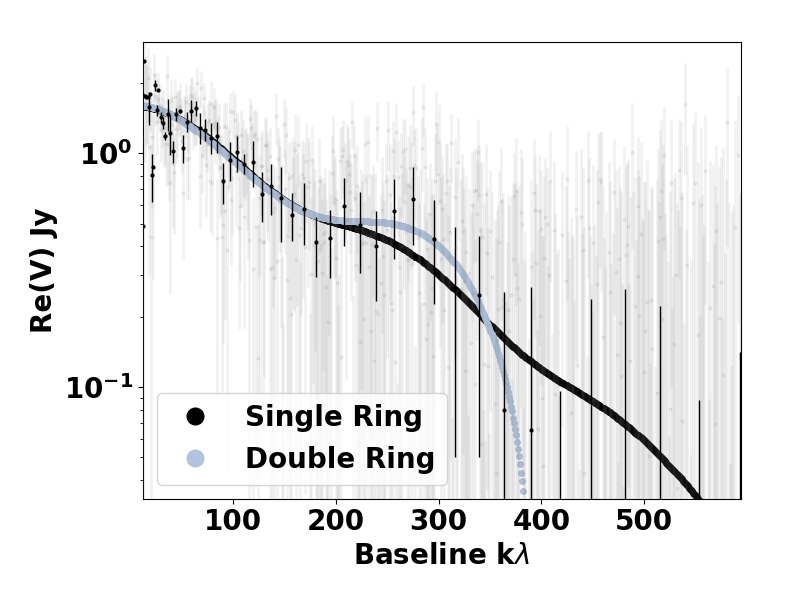}
    \includegraphics[width=\linewidth]{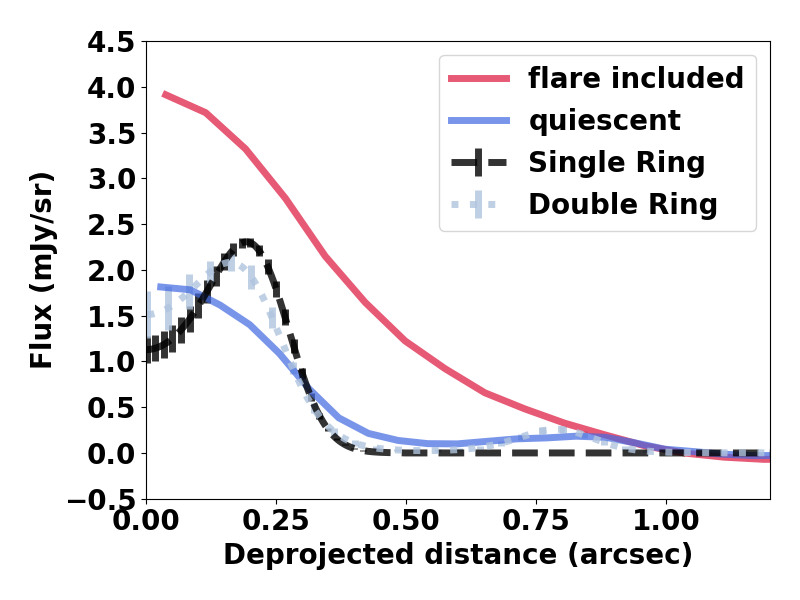}
    
    \caption{\textbf{Top:} A comparison of the two models in Fourier space, where the real component of the visibilities is plotted against baseline length. The coloured lines show the two models, the black points show binned visibilities whilst the grey points are the un-binned visibility points. Here the data being fit is the combination of all observations towards the quiescent disc, i.e. all LB observations and Observation 3. \textbf{Bottom:} Radial flux profiles extracted from continuum images made with and without including flaring observations are shown as solid lines. These are compared to the results from modelling of the observed visibilities. The profiles from visibility modelling are not convolved with any synthesizing beam.} 
    \label{fig:uvspace}
\end{figure}

\begin{figure*}
    \centering
    \includegraphics[width=\linewidth]{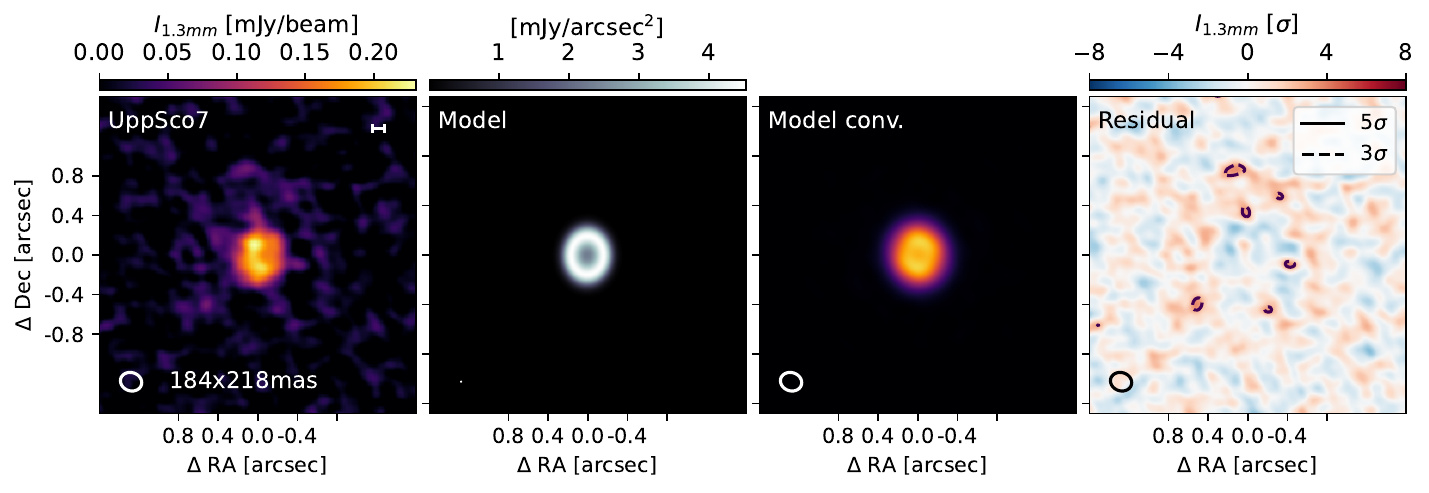}
    \includegraphics[width=\linewidth]{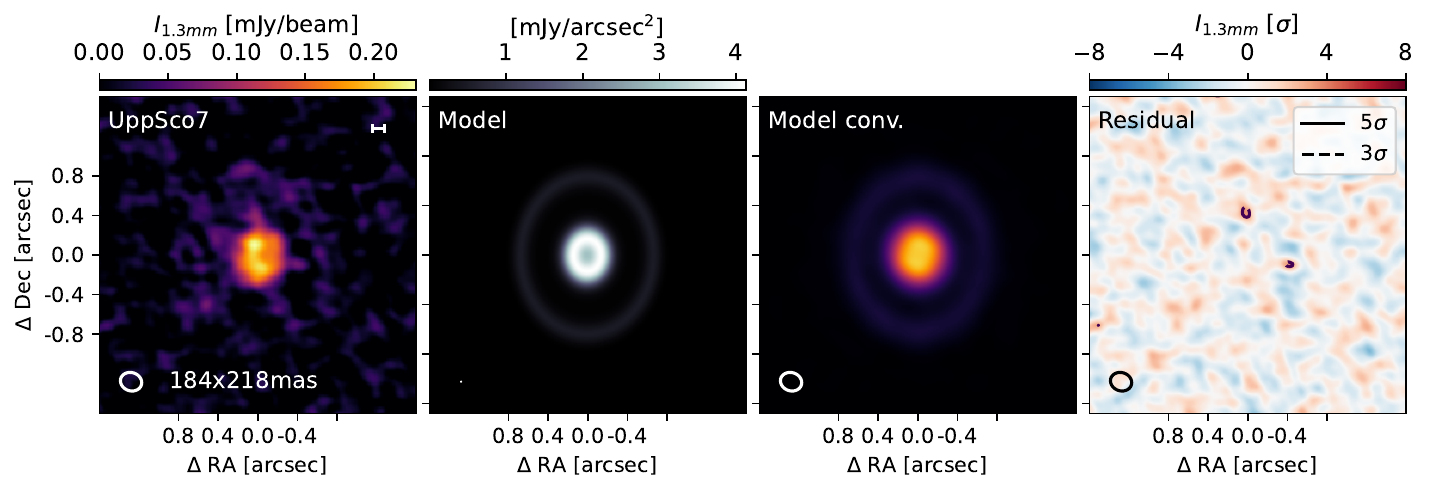}
    \caption{Comparison between the two families of density distributions that result from visibility modelling. On the left of each row the continuum image of the data is shown, here constructed with its native beam size without smoothing or tapering in order to circularise. \textbf{Top Row:} A single ring at short radial separation, \textbf{Bottom row:} A disk comprised of an inner ring and a more extended ring. }
    \label{fig:frank_resids}
\end{figure*}

We use \textsc{frank} \citep{Jennings2020FRANKENSTEIN:Process} to fit non-parametric intensity profiles to the observed visibilities. For details of this fitting procedure see the Appendix, and for an investigation into dust substructures across the full AGE-PRO sample by this method see \citep{Vioque2024TheRadii}. Two models do a very good job of reproducing the observed brightness distribution. One is a double ring model, placing one dust ring at $\approx$ 0$\farcs$2 and one at $\approx$ 0$\farcs$8. The other is a single ring model with a dust ring only at $\approx$ 0$\farcs$2, which can be achieved when fitting a truncated range of baselines that removes the longest baselines that have lower signal-to-noise.
The results are shown as radial intensity profiles and fits to the real component of the visibilities in Figure \ref{fig:uvspace}. In Figure \ref{fig:frank_resids}, we show how the fitted models would look in the image space. We assess the accuracy of the models by subtracting the fitted model visibilities from the observed data, and then imaging the result with CASA. These residual maps are also shown in Figure \ref{fig:frank_resids}.
Both models leave low levels of residuals, as the finer details that distinguish their density distributions are smoothed by the beam of the observation. The double ring model is more successful in representing the faint signal that emerges at greater radial separation from the star, and so here it is marginally preferred. Observations made at shorter, sub-mm wavelengths will be able to illuminate the true nature of this putative outer ring, as they will benefit from the increase in luminosity from smaller grains in comparison to the $\approx$ mm-sized particles traced by these $\lambda=$1.3~mm continuum observations.

\subsection{Probing the nature of the star with analysis of $^{12}$CO (2--1) disk emission}

\begin{figure*}
    \centering
    \includegraphics[width=\textwidth]{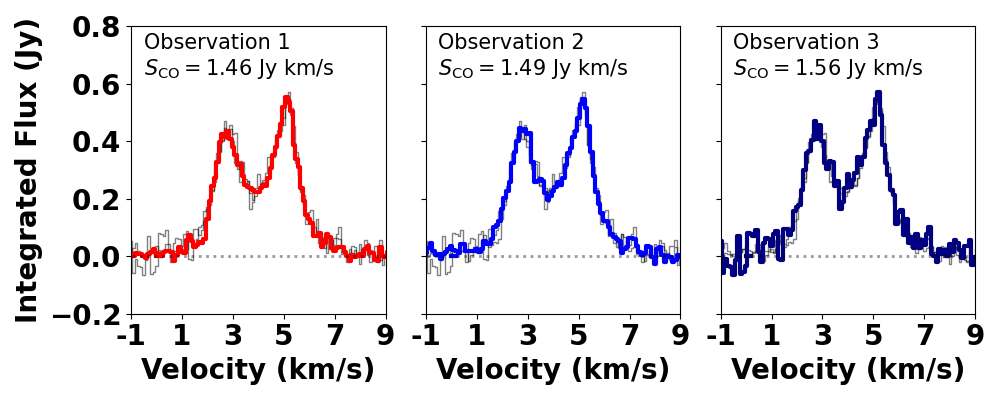}
    \caption{Spectra of the $^{12}$CO emission measured from the compact configuration observations of J16202 extracted using a Keplerian mask. The total integrated flux is given at the top of each panel and the spectra of the particular observation is indicated by a thicker coloured line, thin grey lines show the spectra of the other two observations for direct comparison. }
    \label{fig:flare_gas_ims}
\end{figure*}

Despite a significant variability in the continuum flux, no variability is observed in the measured flux of molecular line transitions. We demonstrate this with the brightest molecular line in J16202 from the AGE-PRO data which is $^{12}$CO J=2--1. Figure \ref{fig:flare_gas_ims} shows spectra from the $^{12}$CO J=2--1 data taken in Observations 1, 2 and 3. To extract the spectra we apply a Keplerian mask using the geometry of the gas disk as given in \citet{Agurto-Gangas2024TheRegion}, and convolve this mask with the synthesizing beam of the image using the package gofish \citep{Teague2019GoFish:Disks}. 

Unlike the continuum flux, the integrated flux from  $^{12}$CO in the disk is consistent within flux calibration uncertainties over the three observations, typically taken to be 10\% for ALMA in Band 6 \citep{ALMAHandbook}. This suggests there is no significant change to the temperature of the gas in emitting regions, for this reason we cannot favour hypotheses where the mm flare is accompanied by disk heating. Unfortunately a reliable detection of HCO+ does not yet exist for J16202, preventing an exploration of potential variable chemistry driven by X-rays \citep[as seen in IM Lup,][]{Cleeves2017VariableChemistry}. 

\begin{figure}
    \centering
    \includegraphics[width=0.9\linewidth]{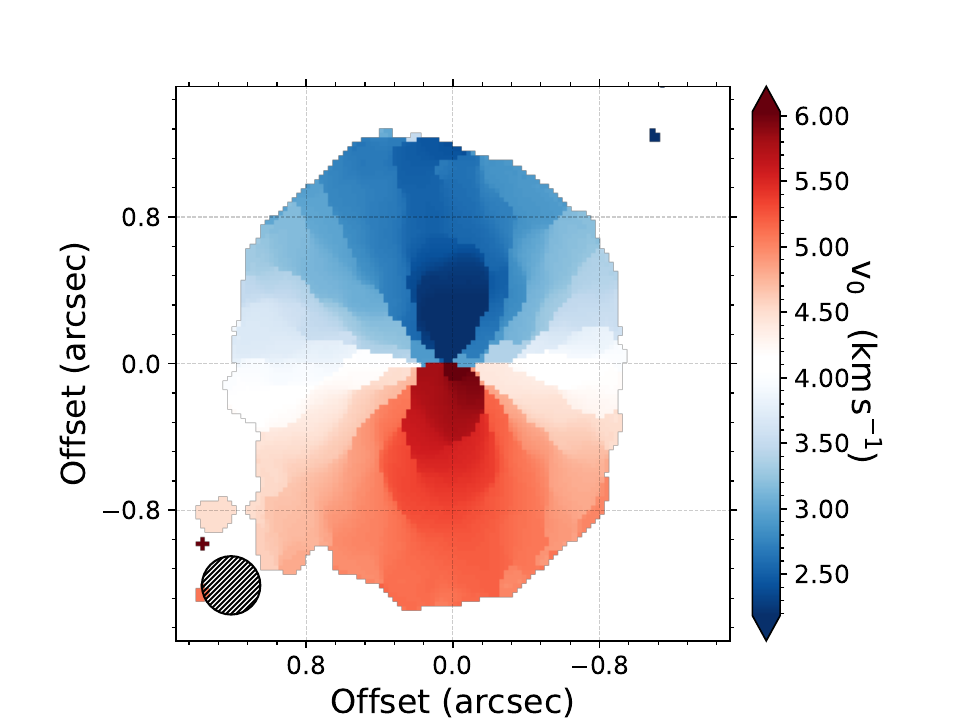}
    
\noindent\includegraphics[width=0.9\linewidth]{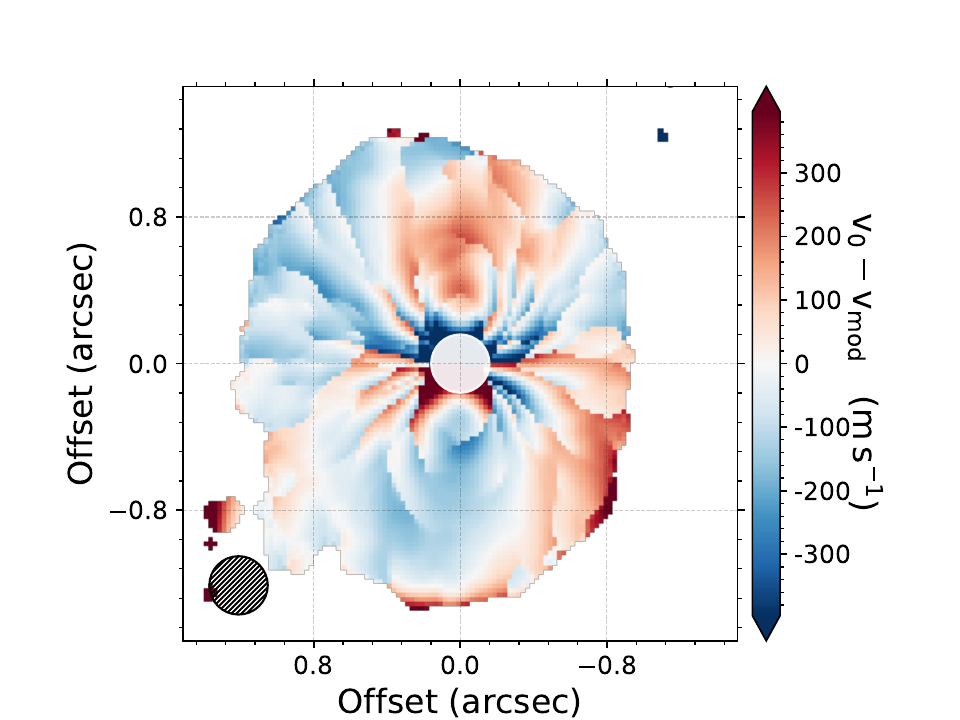}
    \caption{Results of the kinematic analysis of the $^{12}$CO emission detected towards UpperSco7. The top panel shows the velocity field of the $^{12}$CO emission, the bottom panel shows the residual that remains following the subtraction of the fitted Keplerian rotation model. The inner grey ellipse shows one beam area at the centre of the disc, within which results are unconstrained.}
    \label{fig:eddyResults}
\end{figure}

The $^{12}$CO can also help us to characterise the local environment of J16202, which in turn aids to constrain potential flare origin mechanisms. Using \textsc{eddy} \citep{Teague2019EddyDYnamicse} we fit a rotation map to the $^{12}$CO emission cube of J16202. The best fits are achieved when an elevated emission surface is considered, parameterised as $z(r) = z_0~ (r-r_{\rm cav} )^\psi ~e^{ - ( r-r_{\rm cav}/r_{\rm taper} )} $, where $\psi$ relates to the flaring of the emission surface and $z_0$ is related to the aspect ratio of the surface. $r_{\rm cav}$ can be added to consider an inner cavity, as can an exponential taper $r_{\rm taper}$ at the edge of the disc, but our best fits are achieved without this extra complexity in the model. We fix the distance the distance to the Gaia value of 153~pc, and fit for stellar position, the systemic velocity and the stellar mass as well as for $z_0$, $\psi$. The inclination and position angle were fixed to values determined from initial fits that do not use an elevated surface, PA = 179.9 and inclination = 42.3$^\circ$. We ran the fit over all pixels within the cube where emission is detected with at least 3$\sigma$ confidence. Residuals after subtracting the fitted rotation map do not improve if we add an outer disk taper, or by adding a minimum radius to simulate an inner cavity. In Figure \ref{fig:eddyResults} we show the collapse of the image cube using the quadratic method of \citet{Teague2018ALines}, which identifies line centroids, and the residual map following the subtraction of the best fit model. The corner plot showing the results of the Monte Carlo optimisation is presented in Figure \ref{fig:eddy_corner}.

An important result from this procedure is that the stellar mass we derive from fitting a Keplerian rotation map to the observed $^{12}$CO emission of 0.8$\pm$0.1~M$_\odot$ is significantly higher than the mass currently adopted for the star of $0.34^{+0.25}_{-0.47}$~M$_\odot$ \citep[][]{Agurto-Gangas2024TheRegion}. This is a particularly interesting result given our speculation surrounding a binary partner based on the continuum cavity and similar published mm-flaring Class II systems. We note that our fitted inclination value is around 10$^\circ$ greater than that derived from the continuum, where the difference is likely due to the elevated CO surface. If instead we run fits that adopt the geometry achieved by the continuum fitting, PA=179.0 and inclination 32.4$^\circ$, we find a similar level of residuals, but there is less agreement among the walkers and these fits result in an even larger stellar mass of $\sim$ 1.2~M$_\odot$. It could be that the mass of a companion is contributing to this gravitational potential. We note, however, that it is not uncommon for the stellar mass of stars to be underestimated when this is achieved by fitting evolutionary tracks, in particular non-magnetic evolutionary tracks can underestimate M dwarf masses by up to a factor of 2 \citep{Simon2019MassesOphiuchus,Fernandes2023UsingCompleteness}. In this work we will adopt the results as derived from the gas-fitted inclination of 42.2$^\circ$.

Inspecting the residuals in Figure \ref{fig:eddyResults} we see that the azimuthally symmetric Keplerian model describes most of the velocity map in the disk well. The inner disk within $\approx$beam$_{\rm FWHM}$ are less reliably probed, this region is masked with a white ellipse in the residual map of Figure \ref{fig:eddyResults}. The maximum residuals do not reach higher than 0.35 km/s, which is an order of a few times the spectral resolution of the data cube (0.1~km/s). In outer regions of the disk some large scale positive and negative residuals are detected, suggesting that particularly in outer regions, the disk is not circularly Keplerian. Similar positive and negative residuals on large scales are also seen in some circumbinary disks \citep[e.g.][]{Kurtovic2022ThePlanet}. Interestingly \citep{Trapman2024TheMaps} also identify a disk asymmetry but in the $^{12}$CO integrated emission rather than in the velocity field, with strong residuals left following the subtraction of a beam-corrected axisymmetric model. The residuals in this case may be consistent with an unresolved spiral. 

The dynamically inferred mass of 0.8$\pm$0.1~M$_\odot$ would imply a considerable difference in the physical properties of the central object if we are to assume a single host star. The full extent of the range in stellar luminosity and effective temperature that may be inferred from these indicative mass values is demonstrated by Figure \ref{fig:HR} where MIST isochrones calculated by MESA \citep{Choi2016MESAMODELS,Dotter2016MESAIsochrones} are highlighted where they correspond to the age of Upper Sco and to the two stellar mass results under consideration here.
 
\section{Discussion}
\label{sec:discuss}

\subsection{Previously observed millimetre variability in YSOs}
\label{sec:lit}
\subsubsection{Observations of millimetre variability}

Significant variability at mm wavelengths has been identified in only a handful of objects, as shown in the right hand panel of Figure \ref{fig:lit_flares}. In this section we discuss published examples of systems with variable mm emission that can be compared to the case of J16202 in order to establish potential physical origins.  

The data on flaring mm sources are naturally sporadic, as in many cases the discoveries were entirely serendipitous and the observations tuned to serve different scientific goals. As a result, the known flares in mm emission have been measured at various wavelengths and with a variety of cadence in time, either following or during the peak brightness. The light curves in the panels of Figure \ref{fig:lit_flares} demonstrate this wide range of observed decay timescales. The behaviour of mm flux in these systems can vary on scales of days, hours or minutes, as shown by the three left hand panels of Figure \ref{fig:lit_flares}, the lightcurve of J16202 has a comparable timescale to the intermediate and fine timescale panels. The decay timescale of the light curve is most similar to DQ Tau and HD~283572, and is discussed further later in this section. 

The right hand panel of Figure \ref{fig:lit_flares} shows the relative mm flux increase for each of the similar, published cases of mm flaring plotted against a time interval between measurements of peak and quiescent flux in the system.
We note that constraining the timescale and flux increase of such events requires monitoring observations with regular cadence, something currently lacking in the field. Nevertheless, we pursue our analysis with this caveat.
The decay timescales in Figure \ref{fig:lit_flares} are indicative values based on the observational data, some are only provided as upper or lower limits because many of these discoveries were serendipitous or could not sample the full extent of the light curve in each scenario. An additional important consideration when assessing these light curves is that they have not been observed at exactly the same wavelength, but they are all taken at mm or sub-mm wavelengths.

When analysed with finer time resolution the relative increase in flux increases, or a more complex light curve comprised of multiple brightening events on both short and long timescales emerges. Examples can be found in the light curves of HD~283572 \citep{Lovell2024SMA283572} and the Orion Radio Burst Source (named ORBS) reported in \citet{Forbrich2008ARegion,VargasGonzalez2023ACluster}, also known as COUP 647 by \citet{Getman2005CHANDRALISTS}. We note however that \citet{Massi2006SynchrotronA} are able to divide their observations into smaller scan measurements as we do, but variation within individual scans is not significant compared to the longer-term variations. This is also the case for J16202, in Figure \ref{fig:mm_lightcurve} data points are clustered in groups of three measurements made over equal time intervals for each each scan, where the individual scans have an average duration of 85.5s. A better sampling of the light curve could not be achieved in this case due to the need to cycle to other science targets within the observing block to fulfill the original science case of AGE-PRO.

\subsubsection{Comparison of known millimetre-variable objects}

\begin{figure*}
    \centering
    \includegraphics[width=\linewidth]{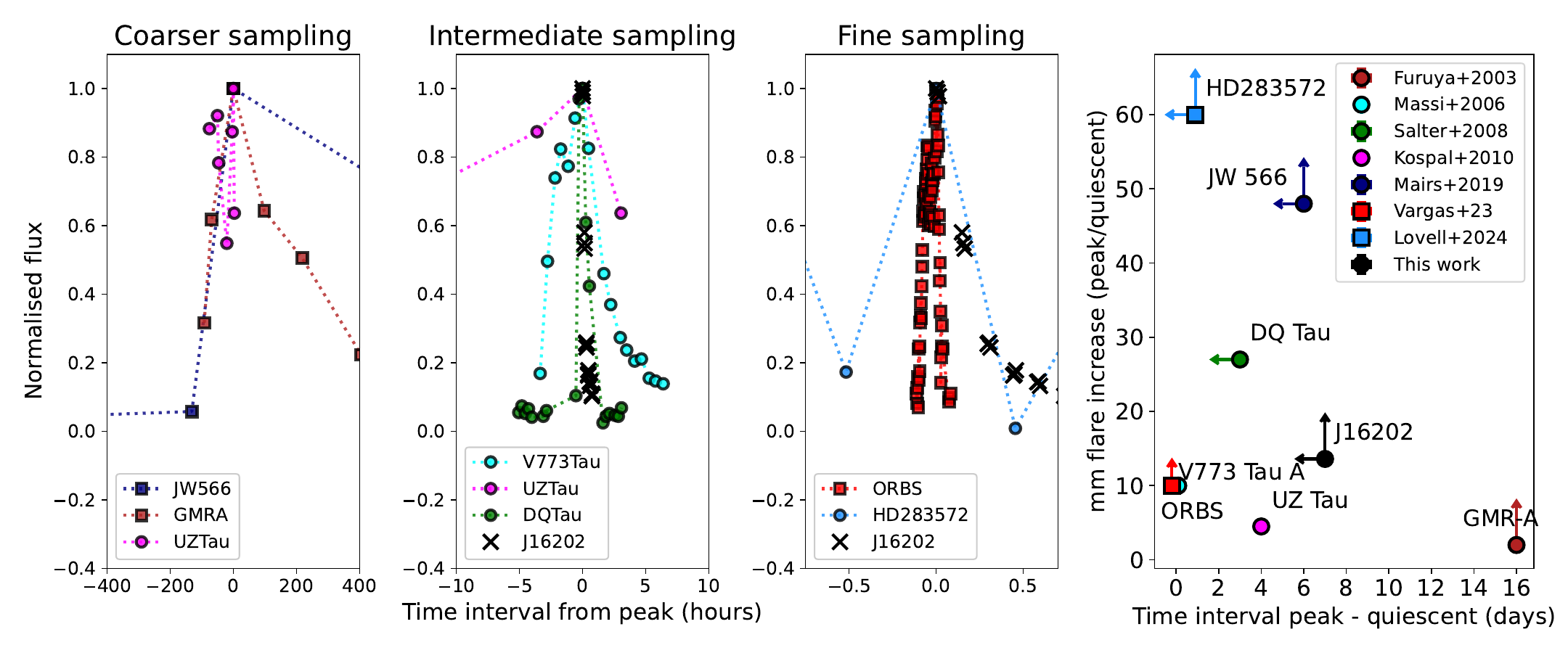}
    \caption{\textbf{Left:} Published millimetre light curves from disk hosting systems with variable flux that exhibit a range of decay timescales. In each plot the measured flux has been normalised to the peak value for each flare. \textbf{Right:} A comparison of the relative flux increase in each system against a time interval measured between the peak brightness and quiescent levels. Where peak flux or duration are unclear in current data, upper and lower limits are used. Circles mark Class II objects, ORBS is classified as Class I, HD~283572 is a Class III object. }\label{fig:lit_flares}
\end{figure*}

The right-hand panel of Figure \ref{fig:lit_flares} shows that mm flare events typically result in an increase of factor $\approx$10 in flux, with the exception of UZ Tau and GMR-A, which increase by a factor of a few. The flares decay on a timescale of $\approx$ days, but this is a highly subjective measurement. For example, V773 Tau A decreases a factor of 7.2 over just a few hours in the observations of the flaring event by \citet{Massi2006SynchrotronA}. When their peak flux is compared to the measurements taken 3 months later, the flux has decreased by factor 120 over the 98 day period. This suggests that the flare behaviour of V773 Tau A is not necessarily limited to the short term, as has also been surmised in radio flaring of YSOs, where short, medium and longer term flaring behaviour is apparent in the sample of \citet{Forbrich2017ExtremeCluster} and in the observations of HD~283572 \citep{Lovell2024SMA283572}. Indeed in UpperSco7 and JW 566, the values in Figure \ref{fig:lit_flares} represent a flare duration defined by a return to detection of quiescent levels, but on shorter time scales a sharper decrease in flux is observed; JW 566's flux decreased by a factor of 2 in just 31 minutes, and J16202 decreases a factor 8.2 in the 50 minutes of Observation 1 (see Figure \ref{fig:mm_lightcurve}).

Ideally, the decay of the flare can be described by an exponential decrease with a characteristic time, but for most of the current sample the time axis not well sampled in order to constrain such a value. Nevertheless, from the current data we can extract indicative, zeroth-order representations of the time-scales over which these phenomena last, and these are presented in Figure \ref{fig:lit_flares}. \citet{Massi2006SynchrotronA} parametrise the decay of the flare light curve in V773 Tau A through an e-folding time, in this case it was presumed to be related to the time scale of energy release from electrons trapped on magnetic fields. In Figure \ref{fig:mm_lightcurve} the best fit exponential curve to the measurements of Observation 1 is shown, which indicates an e-folding time for J16202 of 12 minutes, c.f. 2.3 hours in  V773 Tau A. The implications this timescale might have for indicating a driving physical mechanism are discussed in Section \ref{sec:natureofflare}.


\subsection{Multi-wavelength variability}
\label{subsec:multiwav}

J16202 was previously observed as part of the K2 mission, the optical light curve is presented in \citet{Cody2018The2}. The source is classified as quasi-periodic and symmetric, meaning that the light-curve is neither fading or brightening, with a period of 8.06 days from a periodogram analysis. Extrapolating this 8.06 day period and assuming the periodicity stays constant, there was due to be an optical peak in emission on 30/3/2022, the same day on which the mm flare was observed by ALMA. Observation 3 was taken 7.9 days after Observation 1, just short of the 8.06 day period of the optical variability, and no evidence of flaring emission was detected. As such it remains unclear as to what extent the J16202 mm flare is a periodic phenomena. 

Whilst X-rays might be linked to the mm flares because they share physical mechanisms that lead to the subsequent free-free emission, concurrent optical emission may be telling us more about the dynamics and environment. 
\citet{Salter2008CapturedTauri} note that while DQ Tau also shows periodic optical variability, the magnetospheric interaction that they interpret to be the source of the mm flare cannot also explain the frequency and duration of the optical brightening as well. They speculate that pulsed accretion may occur from the inner binary orbit that disrupts the balance between gas pressure and magnetic pressure at the discs inner edge. This can explain the optical brightening, and the interaction of the binary with the magnetic field could drive mm outflows as well \citep{Basri1997THEPHASE1}. Or indeed vice-versa, where the re-connection events between magnetospheres enable enhanced accretion to drive the optical brightening \citep{Mathieu1997THECURVES,Kospal2011HuntingBinaries}. Simultaneous multi-wavelength monitoring would be the most useful tool in studying these processes further, but such events remain rare; of the 12 binary systems studied by \citet{Kospal2011HuntingBinaries} only UZ Tau showed variability in both mm and optical measurements. 
             
Some of the other known mm flaring systems also exhibit bright X-ray emission or optical emission alongside the mm flare. To what extent these optical or X-ray emissions are associated with mm flares is not entirely clear. In some cases flares are perfectly coincident \citep[e.g. optical and mm in UZ Tau ][]{Kospal2011HuntingBinaries}, in others the flares do not match well at all (e.g. 2mm, 3mm, and K band behave differently in observations of GMR-A \citep{Furuya2003AWavelengths}).

X-ray flaring is more common than mm flaring, with X-ray flare mechanisms potentially linked to the non-thermal origin of mm and radio flux. \citet{Guedel1993X-Ray/MicrowaveS} report a correlation between quiescent radio and X-ray luminosities across magnetically active stars, though the underlying relationship remains complex. Using VLA and Chandra data—sometimes simultaneously—\citet{Forbrich2017ExtremeCluster} identify 13 YSOs in the Orion Nebula Cluster with radio variability exceeding an order of magnitude, all of which are X-ray sources. However, only 5 of the 18 most variable X-ray sources have radio counterparts, and just two of these overlap with the extreme radio variability sample. The authors conclude that extreme radio and X-ray variability correlate only on short timescales ($<$1 day).

The connection between mm variability and X-ray emission is particularly relevant for Class II discs due to potential impacts on disk chemistry. For example, an increase in H$^{13}$CO+ emission in IM Lup is interpreted by \citet{Cleeves2017VariableChemistry} as evidence of X-ray-driven chemistry. While time-variable chemistry in protoplanetary discs remains underexplored, current studies suggest it may significantly influence certain molecules critical to planet formation \citep{Cleeves2017VariableChemistry, Waggoner2022ClassificationDisks}.

\subsection{The nature of the flare in J16202}
\label{sec:natureofflare}
What physical mechanism could have driven the dramatic mm flare that we observe? We can already speculate on likely origins on the basis of the data already in hand. The short-lived bright emission observed towards J16202 in our observations is not likely to be purely thermal emission, as the spectral indices calculated from the observations in Table \ref{tab:specInd} give a clear indication that quiescent phases are consistent with thermal emission from optically thin dust, whilst most of the flaring phases must be dominated by non-thermal emission.

The very quick decay of the emission throughout the duration of Observation 1 (Figure \ref{fig:mm_lightcurve}) is also consistent with a non-thermal origin. In this scenario electrons trapped in magnetic field loops will spiral along field lines, releasing energy via magneto-Brehmstrahlung until their kinetic energy is spent. 
We assume that the magnetic fields originate from a star, or stars, in the system rather than the field associated with the protoplanetary discs, which is of much lower magnetic field strength and mainly drives turbulence within the disc. In this scenario the timescale over which emission exceeds that of quiescent levels is related to the time that electrons spend trapped along magnetic fields lines radiating their energy, unless multiple flare events are indeed occurring \citep[e.g.][]{Massi2002PeriodicTauri}.
The gyro-frequency of the emission can be related to the strength of the magnetic field that accelerates the electrons with the following expression \citep[see][]{Gudel2002StellarGiants};

\begin{equation}
    \nu = s \nu_{\rm c, rel} \approx 2.8\times 10^{6} B \gamma^2 ,
\end{equation}

where $s$ is the harmonic number, $\nu_{\rm c, rel}$ is the relativistic cyclotron frequency, $B$ the magnetic field strength and $\gamma$ is the Lorenz factor. \citet{Shulyak2019MagneticSurvey} measure magnetic fields in M dwarfs to be of order kG, assuming $\nu $ to be the frequency of our ALMA observations, this would require higher harmonic numbers that are most typical of synchrotron emission from relativistic electrons, rather than the gyro-synchrotron process originating from mildly relativistic electrons.

The specific magnetic structure that contains the emitting particles remains unclear from the current data.
We have reviewed the small number of comparable phenomena in the published literature in Section \ref{sec:lit}, the Class II discs DQ Tau and V773 Tau A are the most analogous examples for the case explored here of J16202. Both systems are also pre-main sequence T Tauri systems with circumbinary discs. Both of these systems are close binaries with eccentric orbits, which enables the scenario whereby magnetospheric connection is made between the fields of each of the binary companions as they approach periastron. In DQ Tau, with period of 15.6 days, the periastron approach of the highly eccentric orbit results in a very close pass of $\lesssim 2R_*$ where the magnetospheres interact. There can also be interactions between magnetic fields at slightly greater separations. In the case of V773 Tau A \citep{Salter2008CapturedTauri}, extended coronal features similar to solar helmet streamers bridge a larger separation to cause the interaction \citep{Massi2008InteractingStreamers}.
A dynamically cleared inner disk leads to reduced magnetic braking of the stellar rotation from star-disk field lines \citep{Basri1997THEPHASE1}, which might allow field lines to extend further outwards. In this system the binary period of 51 days is equal to the periodicity of the mm brightness variations \citep{Massi2006SynchrotronA}.

In J16202 there is evidence of an inner cavity to the dust disk (Figure \ref{fig:frank_resids}) and tentative evidence of gas depletion in the inner disk from the modelling of \citet{Trapman2024TheMaps}. The peak brightness in the single ring model is at a separation of 0$\farcs$19 ($\approx$30 au), and 0$\farcs$16 ($\approx$25 au) for the double ring model. If driven only by the dynamics of an inner orbiting binary, the size of the cavity might indicate binary orbit properties. Although we note that that a companion with an orbit matching the timescale of K2 variability (8 days) would not be capable of opening a gap this large. Indeed some circumbinary systems such as AK Sco, V4046 Sgr  have cavities which are much larger than the predicted size of a cavity should it be driven solely by their known companions. \citet{Tokovinin2019BinaryScorpius} previously observed J16202 using speckle interferometry observations, with which they are able to rule out a binary companion to J16202 down to a separation of 64 mas, equivalent to $\approx$9.8 au at the distance of the system, consistent with what is inferred from the ALMA imaging. Our best fitting model of the visibilities favours a dust distribution that includes an inner cavity, although high resolution observations of mm dust and molecular gas distributions are key to confirming this.

It is tempting to draw a link between a supposed binary orbit and the quasi-periodic symmetric light curve from Kepler observations of the system. If the mm and optical behaviour originate from the same close passage at binary periastron, then we would expect both to have similar periodic behaviour. The quasi-periodic symmetric sources have similar timescales to dipper disks in which occultation of light by circumstellar material is the presumed cause of optical variability \citep{Ansdell2016YOUNGK2}. The quasi-periodic symmetric category of \citet{Cody2018The2} spans a broad range in parameter space, and may be comprised of a number of sub-categories. For example their variability at optical wavelengths may be related to the presence of starspots \citep{Cody2018The2}. Overall quasi-periodic symmetric sources generally show greater amplitudes and strong H$\alpha$ emission, which could signal stronger gas accretion. 

Analysis of the $^{12}$CO emission cube provides further circumstantial evidence that adds to the speculation surrounding binarity, due to the large dynamical mass calculated and the large-scale residuals following the subtraction of a Keplerian rotation map. In this case, the scenario of binary interaction as a cause for the mm flare seems most likely. However, perturbations to the outer regions of protoplanetary discs do not require stellar binarity, for example misalignments in discs can also be driven by giant planets hosted within the system, \citep[see e.g.][]{Nealon2015ApsidalDiscs, Nealon2018WarpingOrbit}. In this case a single star scenario is still possible, and the mm flaring can be interpreted as activity at the stellar surface of the star. 
However, we note the results of Section \ref{sec:results}, that the innermost regions on scales comparable to the synthesised beam are not well constrained by the current data. 

Future high resolution images of the inner disk of J16202 will be key in truly distinguishing these scenarios. Robust characterisation of the system also requires a better understanding of the central object in J16202. Establishing whether or not a binary partner exists will immediately determine the validity of individual proposed mechanisms, for example is the separation short enough to enable helmet streamers as seen in V773 Tau A \citep{Massi2008InteractingStreamers}.  Such data also will focus the scope of future follow up observations. A companion of equal mass to the primary, orbiting with a period of 8.06 days and assuming an inclination the same as that of the protoplanetary disk would results in an RV amplitude of $\approx$34~km/s, or higher in the case of an eccentric orbit. This can be easily detected by many modern RV instruments.

Another potential origin mechanism to consider is a powerful accretion burst. An increase in mass transport through the disk could also result in a short-duration, discrete spike in emission such as we observe. 
In low mass stars, magnetospheric accretion is the preferred accretion mechanism. An increase in magnetospheric accretion can result in an increase of free-free emission. This signal would originate from a very small region at the stellar position, resulting in an un-resolved emitting source with a negative spectral index, meaning it is broadly consistent with the basic properties of observed mm flare. The exact nature of the mechanism that drives the burst in emission from this accretion is unclear. Attempting to model the magnitude or duration of such an event is beyond the scope of the current work, it requires knowledge of a number of properties about the exact accretion mechanism and the inner disk environment of the system that currently remain entirely unconstrained. For example we currently lack observations that can help to constrain the accretion behaviour in J16202. Future observations of accretion tracing lies (e.g. H$\alpha$ , He I) could be used to determine the accretion rate in J16202. Other Upper Sco targets from the AGE-PRO sample show accretion rates over a range -10.98  $< log\dot M_{\odot} (yr^{-1}) <$-8.8 \citep{Agurto-Gangas2024TheRegion}, offering indicative values for a Class II system in Upper Sco. More generally, \citet{Manara2020X-shooterMyr} find Upper Sco accretion rates to be broadly similar to those in Lupus and Cha I, noting that even those with low disk masses can have high accretion rates. Repeated observations of accretion tracers in J16202 at multiple epochs would help to constrain the variability of this accretion and facilitate a comparison with other Class II discs both in Upper Sco and beyond. Nevertheless this remains a possible mechanism for creating free-free emission on unresolved spatial scales in this system. 
Following the initial accretion event, the star will increase in luminosity which subsequently heats the circumstellar disk. This too will increase the mm emission from the disc, but to a much lesser magnitude and likely over a more extended timescale that will depend upon complex radiative transfer processes; in order to emit at mm wavelengths, the heating would have to penetrate to the dense, cool midplane of the disc. It would be unlikely for there to be a sufficient disk temperature increase in order to reproduce the measured order-of-magnitude flux increase in the short timescale of the observed flare in Observation 1. However, it might be able to explain longer term, more subtle variations in emission, such as was observed in Observation 2 and discussed in the context of the `active quiescent' phases of HD~283527. If this were the case we would expect for there to be a concurrent increase in gas temperature as well, but in Figure \ref{fig:flare_gas_ims} we see no evidence of such an occurrence.

\section{Conclusions}

We present ALMA observations that detect an extraordinary flare in mm emission towards J16202. The first observation catches the tail of the flare with a peak initial flux of 20.5 mJy, before fading by a factor of $\approx$8.5 over 50 minutes. Just under 8 days later, the disk has reached what we presume to be quiescent levels of $\approx$ 1.5 mJy.

The rapid dimming and intra-band spectral index calculated during the flare observations implies a non-thermal origin to the flare emission. The properties of the flare emission are consistent with synchrotron emission. The observed behaviour is similar to cases of the collision of the magnetospheres belonging to binary companions. The re-connection event can occur through streamers related to extended magnetic structures, or in the case of an eccentric binary, direct overlap of the two fields at periastron. There is not yet evidence of binarity in the J16202 system, but a close binary companion could be hidden at separations smaller than 10~au that have not been probed by existing observations. Alternatively, the flare may come from the interaction of magnetic loops at an active stellar surface with magnetic field strength of order $\approx$ kG. Other potential explanations might include the interaction of magnetic loops at an active stellar surface, or a massive accretion burst if the host is in fact a single star.

More observational data on flaring mm systems is sorely needed in order to investigate these phenomena further. We leave here some suggestions that future observing campaigns might consider.
Firstly, flares at millimetre wavelengths must be observed with a much finer sampling cadence, this could be achieved through dedicated monitoring programs or observations that follow-up a flare event. This will allow for better characterisation of the light curve and tighter constraints on measured decay times. If monitored for a sufficient period of time, these observations will also be able to constrain whether or not flares occur periodically, and whether a single peak in brightness or multiple peaks are present over the course of the period. These are key properties that will distinguish between the candidate structures for magnetic connection events in YSOs. Future observing campaigns should aim to coordinate simultaneous observations at higher frequencies, in particular monitoring for concurrent X-ray or optical variability. 

To investigate the impact of such events on the circumstellar disc, simultaneous observations of key radical transitions related to variable chemistry (e.g. HCO+) should be scheduled during flaring events. The non-detection in the case of J16202 suggests that such observations will have to be highly sensitive. In relation to UpperSco 7 specifically, spectroscopic or VLTI observations could definitively rule in or out the possibility of a central binary and the potential magnetic connection mechanisms associated with close binarity.

\section*{Acknowledgments}

Accepted for publication in ApJ for the special issue of AGE-PRO.

This paper makes use of the following ALMA data: ADS/JAO.ALMA\#2021.1.00128.L, ALMA is a partnership of ESO (representing its member states), NSF (USA) and NINS (Japan), together with NRC (Canada), MOST and ASIAA (Taiwan), and KASI (Republic of Korea), in cooperation with the Republic of Chile. The Joint ALMA Observatory is operated by ESO, AUI/NRAO and NAOJ. The National Radio Astronomy Observatory is a facility of the National Science Foundation operated under cooperative agreement by Associated Universities, Inc.

J.M. acknowledges support from FONDECYT de Postdoctorado 2024 \#3240612 and from the Millennium Nucleus on Young Exoplanets and their Moons (YEMS), ANID - Center Code  NCN2024\_001.
L.P. acknowledges support from ANID BASAL project FB210003 and ANID FONDECYT Regular \#1221442
C.A.G. acknowledges support from FONDECYT de Postdoctorado 2021 \#3210520
A.S. acknowledges support from FONDECYT de Postdoctorado 2022 \#3220495 and support from the UK Research and Innovation (UKRI) under the UK government’s Horizon Europe funding guarantee from ERC (under grant agreement No 101076489)
L.T. acknowledges the support of NSF AAG grant \#2205617
R.A. acknowledges funding from the Fondazione Cariplo, grant no. 2022-1217, and the European Research Council (ERC) under the European Union’s Horizon Europe Research \& Innovation Programme under grant agreement no. 101039651 (DiscEvol). Views and opinions expressed are however those of the author(s) only, and do not necessarily reflect those of the European Union or the European Research Council Executive Agency. Neither the European Union nor the granting authority can be held responsible for them.
P.P. acknowledges the support from the UK Research and Innovation (UKRI) under the UK government’s Horizon Europe funding guarantee from ERC (under grant agreement No 101076489)
I.P. and D.D. acknowledge support from Collaborative NSF Astronomy \& Astrophysics Research grant (ID: 2205870)
L.A.C. acknowledges support from the Millennium Nucleus on Young Exoplanets and their Moons (YEMS), ANID - Center Code   NCN2024\_001 and the FONDECYT grant \#1241056.
K.Z. acknowledges the support of the NSF AAG grant \#2205617
C.G-R. acknowledges support from the Millennium Nucleus on Young Exoplanets and their Moons (YEMS), ANID - Center Code  NCN2024\_001
G.R. acknowledges funding from the Fondazione Cariplo, grant no. 2022-1217, and the European Research Council (ERC) under the European Union’s Horizon Europe Research \& Innovation Programme under grant agreement no. 101039651 (DiscEvol). Views and opinions expressed are however those of the author(s) only, and do not necessarily reflect those of the European Union or the European Research Council Executive Agency. Neither the European Union nor the granting authority can be held responsible for them.
The authors would like to thank colleagues for useful and insightful discussions on topics covered in this article, with particular thanks due for advice from Josh Lovell on analysing mm lightcurves, from Sebastian Perez on the nature of accretion bursts and their emission processes and from Jaime Vargas on mm variability in YSOs and systematic searches to find them.


\bibliography{references}

\bibliographystyle{aasjournal}

\appendix

\section{Continuum visibility fits with \textsc{frank}}

For visibility modelling of the entire AGE-PRO sample, and detailed description of the fitting procedure, see \citet{Vioque2024TheRadii}. Here we model only the non-flaring data, and we include the long-baseline observations in order to be sensitive to the shortest possible length scales . \textsc{frank} fits are controlled by hyperparameters that dictate the SNR threshold for visibilities to be fitted and the extent to which the power spectrum estimate is smoothed \citep{Jennings2020FRANKENSTEIN:Process}. We make fits that vary these hyperparameters as well as the outer spatial scales to which the data should be fit. 
From this exploration, two types of dust distribution emerge frequently; one where there is single ring, and one with a double ring which attempts to describe very faint and extended outer-disk emission. These are achieved using frank hyperparameters $\alpha=$1.2, w$_{\rm smooth}$=10$^{-2}$, R$_{\rm max}$ = 2.0$''$ for a single ring result and $\alpha=$1.05, w$_{\rm smooth}$=10$^{-4}$, R$_{\rm max}$ = 2.0$''$ for the double ring result. In Figure \ref{fig:uvspace} examples of these two fits are shown. When the fit is restricted to a relatively short radius from the stellar position, e.g. R$_{\rm max}$ = 0.5$''$, compact discs with no rings or substructure result, but these are a bad fit to the deprojected visibilities. When the radius is extended, rings emerge from the fitting process.

\section{Corner plots of the fit to a Keplerian rotation model}

\begin{figure*}[h]
\centering
\includegraphics[width=\linewidth]{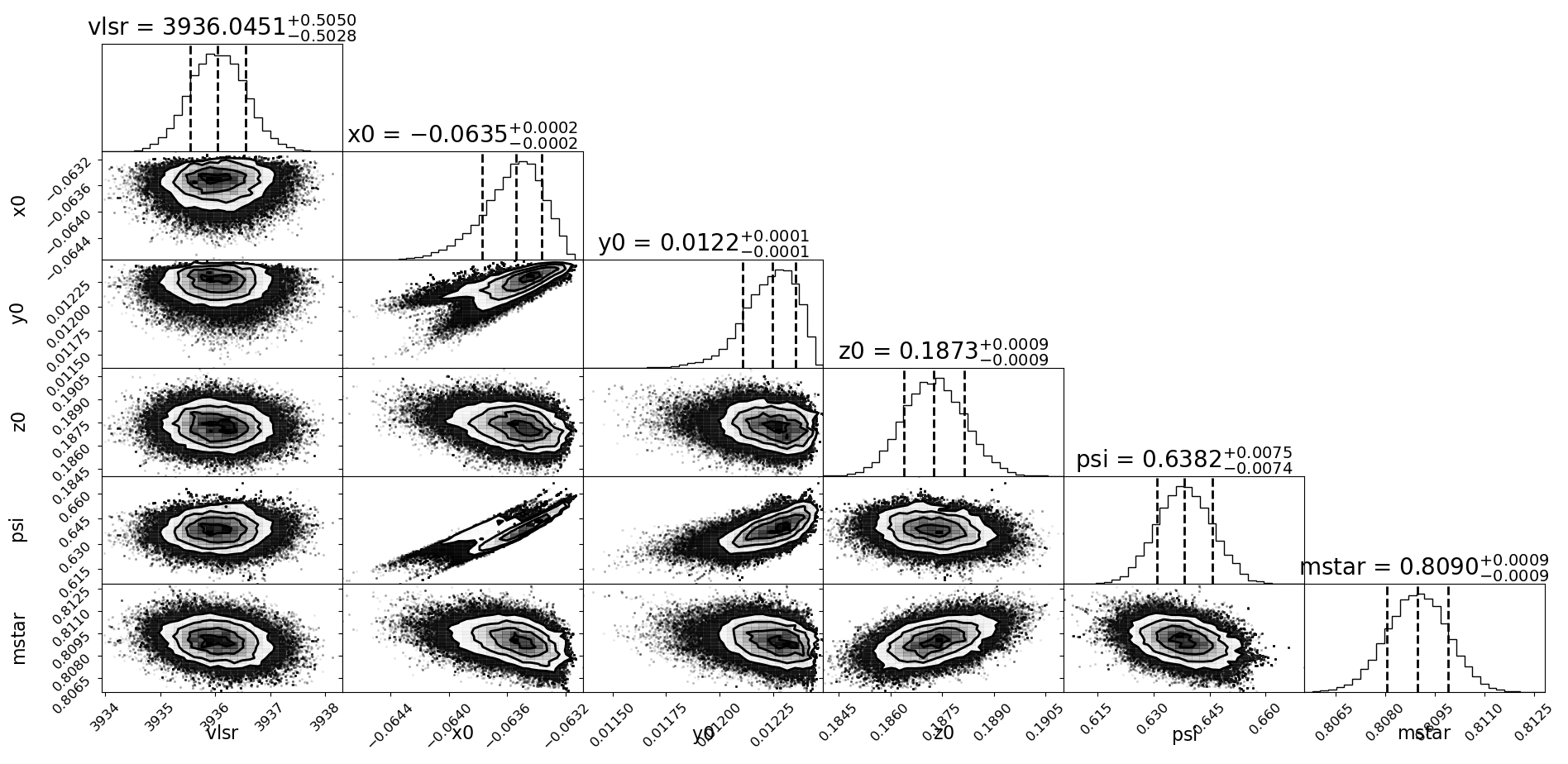}
\caption{Corner plot of the Markov Chain Monte Carlo (MCMC) fitting of a Keplerian rotation map to the $^{12}$CO J=2--1 velocity field. The MCMC was run for a a total of 5 iterations, each using 2000 steps, with 200 additional burn-in steps.}
\label{fig:eddy_corner}
\end{figure*}

Figure \ref{fig:eddy_corner} shows the projections of the posterior probability distributions of the parameters in the rotation map fit to the observed $^{12}$CO line emission from the disk of J16202. 

\section{Hertzsprung Russel diagram comparing stellar properties inferred from mass estimates}

\begin{figure}
    \centering
    \includegraphics[width=0.5\linewidth]{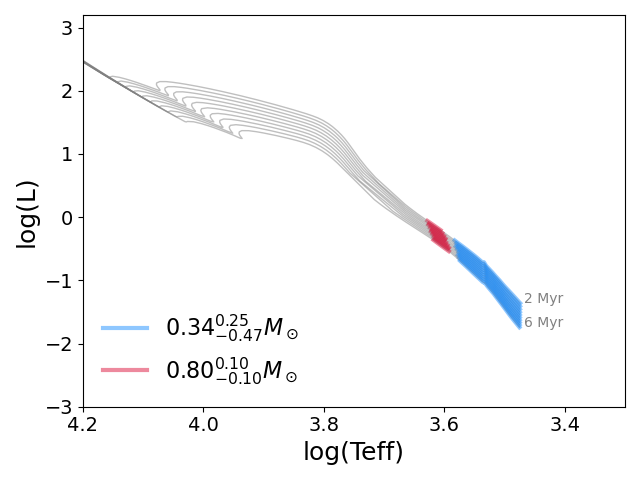}
    \caption{Hertzsprung Russell diagram showing isochrones within the age range of Upper Sco sources (taken to be 2-6 Myr) and assuming solar metallicity. We highlight models with a stellar mass that is equivalent, within uncertainties, to the spectroscopically determined value of 0.34 M$_\odot$ and the dynamically derived result from this work of 0.8 M$_\odot$.}
    \label{fig:HR}
\end{figure}

In Figure \ref{fig:HR} we identify MIST isochrones between the ages of 2-6 Myr (i.e. corresponding to the age of Upper Sco in our AGE-PRO sample) that are consistent with the stellar mass value derived from spectroscopic characterisation and with the dynamically inferred stellar mass achieved in this work (Fig \ref{fig:eddyResults}). The uncertainties on the emcee fit to the rotation map as presented in Figure \ref{fig:eddy_corner} represent a statistical fit to the data, but they underestimate the true uncertainty in the mass of the star that results from this procedure. In this work we take a more conservative uncertainty range of 0.8$\pm0.1$ M$_\odot$. Dynamical masses estimated with  Figure \ref{fig:HR} demonstrates the full range in luminosity and effective temperature values that current mass estimates would infer, whilst remaining consistent to simulations of stellar evolution. We present here an initial kinematic analysis of the rotation of $^{12}$CO in the molecular gas disk around J16202, future work will undertake an in-depth modelling of dynamically derived masses for the full AGE-PRO sample.

\end{document}